\documentclass[aip,reprint]{revtex4-1}
\usepackage{graphicx}
\usepackage[T1]{fontenc}
\usepackage{amsmath}

\draft 


\begin{document}

%\begin{table*}[t!]
%\begin{minipage}{1\textwidth}
%\begin{flushleft}
%This article may be downloaded for personal use only.\\ Any other use requires prior permission of the author and AIP Publishing. \\The following article appeared in Applied Physics Letters and may be found at \url{https://doi.org/10.1063/1.5110414}.
%\end{flushleft}
%\end{minipage}
%\end{table*}

\title{Three-dimensional charge transport mapping by two-photon absorption edge transient-current technique in synthetic single-crystalline diamond}

\author{C. Dorfer}
\email[Author to whom correspondence should be addressed. Electronic mail: ]{dorfer@phys.ethz.ch}
\affiliation{Institute for Particle Physics and Astrophysics, ETH Zurich, 8093 Zurich, Switzerland}
\author{D. Hits}
\affiliation{Institute for Particle Physics and Astrophysics, ETH Zurich, 8093 Zurich, Switzerland}
\author{L. Kasmi}
\affiliation{Institute of Quantum Electronics, ETH Zurich, 8093 Zurich, Switzerland}
\author{G. Kramberger}
\affiliation{Department of Experimental Particle Physics, Jo\v{z}ef Stefan Institute and Department of Physics, University of Ljubljana, 1000 Ljubljana, Slovenia}
\author{M. Lucchini}
\affiliation{Dipartimento di Fisica, Politecnico Milano, 20133 Milano, Italy}
\author{M. Miku\v{z}}
\affiliation{Department of Experimental Particle Physics, Jo\v{z}ef Stefan Institute and Department of Physics, University of Ljubljana, 1000 Ljubljana, Slovenia}
\author{R. Wallny}
\affiliation{Institute for Particle Physics and Astrophysics, ETH Zurich, 8093 Zurich, Switzerland}

\date{\today}

\begin{abstract}
\noindent We demonstrate the application of two-photon absorption transient current technique to wide bandgap semiconductors. We utilize it to probe charge transport properties of single-crystal Chemical Vapor Deposition (scCVD) diamond. The charge carriers, inside the scCVD diamond sample, are excited by a femtosecond laser through simultaneous absorption of two photons. Due to the nature of two-photon absorption, the generation of charge carriers is confined in space (3-D) around the focal point of the laser. Such localized charge injection allows to probe the charge transport properties of the semiconductor bulk with a fine-grained 3-D resolution. Exploiting spatial confinement of the generated charge, the electrical field of the diamond bulk was mapped at different depths and compared to an X-ray diffraction topograph of the sample. Measurements utilizing this method provide a unique way of exploring spatial variations of charge transport properties in transparent wide-bandgap semiconductors.
\end{abstract}

\maketitle

The development of electronic devices based on wide-bandgap semiconductors, as for example, for power electronics\cite{green_diamond} or charged particle detectors \cite{Lukas_PSD}, requires a detailed knowledge of the charge carrier transport properties in those materials. In this letter we present an extension of the known transient current technique (TCT) to obtain such knowledge in wide bandgap materials. It is based on  generating charge carriers in the bulk of a wide-bandgap semiconductor through two-photon absorption (TPA) of a femtosecond laser pulse. The transient current pulse formed by the movement of the carriers in the electric field of the bulk is then observed and analyzed. 

\indent The laser beam enters through the edge of the sample, thus giving the name to the method: edge-TCT. The ability to generate, within femtoseconds, spatially confined charge carriers in any point within the specimen accessible by a laser beam, allows to disentangle the influence of local variations in the electric field from trapping of free charge carriers. This gives TPA edge-TCT a significant advantage over the traditional TCT methods, where the carrier are either confined in space at the surface (top-TCT) or distributed along the beam line (edge-TCT). Among the traditional methods are above-bandgap laser excitation\cite{gabrysch} or the use of alpha particles. Both types of signal generation do not allow for a 3-dimensional scanning of the bulk. In addition, the absence of independent knowledge of the carrier generation time in the case of alpha particles further complicates the analysis.

\indent The edge-TCT method was first developed for silicon devices for charged particle tracking\cite{opa_kramberger2010}. It uses a laser beam with sub-bandgap wavelength that excites the carriers through the device edge along a narrow ($\sim$10\,$\rm \mu m$) path. The current pulse evolution is subsequently observed on one of the narrow strip electrodes perpendicular to the laser path. Recently the TPA edge-TCT has also been demonstrated in silicon devices \cite{tpa2,tpa3,tpa4}. While in silicon devices the TPA edge-TCT requires a more complex laser setup in comparison to the traditional sub-bandgap edge-TCT, in wide-bandgap semiconductors the TPA edge-TCT is the only feasible method. This is because the generation and fine tuning of a deep ultra-violet (DUV) laser beam with a wavelength slightly below the bandgap energy, for example in diamond samples, presents significant difficulties. It requires relatively complex laser sources such as DUV dye lasers or multistage frequency doubling with an optical parametric amplifier, which further increases the cost and complicates the setup. In addition, charge in sub-bandgap wavelength experiments is generated all along the light path and not spatially confined around the focal point as in TPA edge-TCT.

\begin{figure}[ht]
  \centering
    \includegraphics[width=0.47\textwidth]{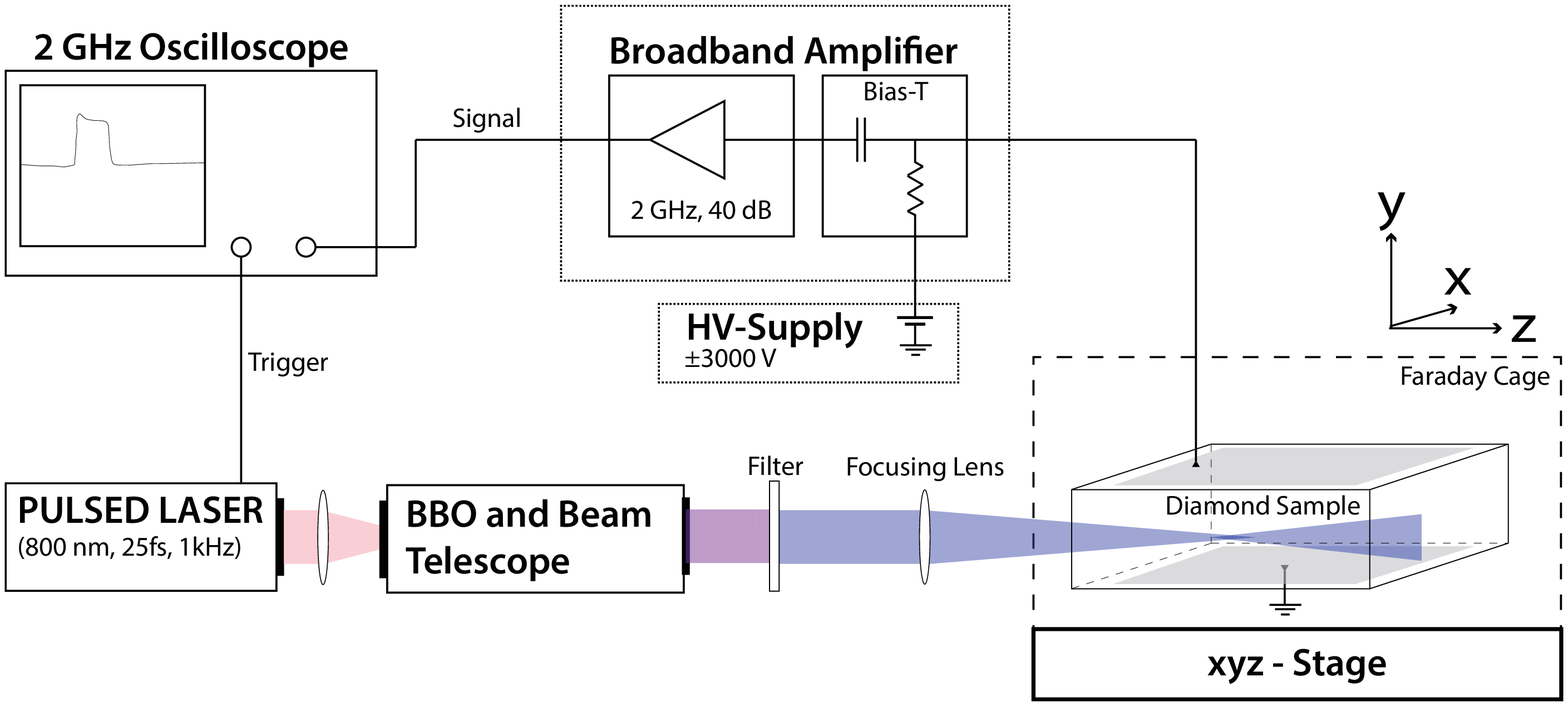}
  \caption{Schematic view of experimental setup}
  \label{pic:sensor}
\end{figure}

\indent Figure~\ref{pic:sensor} shows a simplified overview of our setup. The Ti:Sapphire laser system (Spectra-Physics Femtopower Pro V CEP) of the ETH Ultrafast Laser Physics group\cite{uflgroup,lucchini_laser} provides 25-fs laser pulses with 800-nm wavelength and pulse energies of up to 2.2\,mJ at a repetition rate of 1\,kHz. Only a sub-percent reflection of the initial beam is sent to the setup in which the second harmonic (400\,nm) was generated by a 100-$\rm \mu m$ thick barium borate (BBO) crystal. The beam is then sent through a filter to remove residual IR components from the 400-nm wavelength light. Focusing the beam with a $f$=30\,mm achromatic doublet along the z-axis into the diamond results in a focal point with full width at half maximum (FWHM) of 1.6\,$\rm \mu m$ and Rayleigh length of 28\,$\rm \mu m$ as measured with the "knife edge" technique in air. Typical pulse energies for the experiment ranged from 0.1-0.2\,nJ with pulse intensity stable to below 1\%. Due to the high refractive index of diamond, refraction and dispersive pulse broadening should be taken into account. Refraction causes the focal point to elongate and shifts its position forward in the direction of the beam when entering the diamond. While the diameter remains approximately the same, the Rayleigh length increases by roughly the refractive index of diamond. The z-position of the focal point inside diamond is affected in a similar way. Therefore, in order to determine the actual z-position inside the diamond, the z-displacement of the xyz-stage was corrected by the index of refraction and referenced to the sample edge through which the laser pulse enters. Dispersive pulse broadening causes an increase of pulse length from 25\,fs to roughly 100\,fs after passing through the optical elements and {$\sim$2\,mm} of diamond. Based on observed laser power dependence of the signal, we conclude that two-photon absorption (TPA) is the prevalent mechanism of charge generation in diamond, which agrees with published absorption constants~\cite{Kozak12}.

\indent The diamond under test was purchased from Element6\footnote{Element Six Technologies Limited, Kings Ride Park, Ascot, Berkshire, SL5 8BP, United Kingdom}. It was a 4660$\times$4790$\times$540\,$\mu m^3$ scCVD diamond with the large surfaces oriented perpendicular to [001] crystallographic orientation. A 4100$\times$4100\,$\mu m^2$ Ti:W square pad was sputtered to both larger surfaces through shadow mask sputtering after which the sample was annealed at 400\,$^\circ$C for 5 minutes in N$_{\text{2}}$ atmosphere. In order to reduce dispersion at the edges, two of the diamond smaller sides were polished. Since the injected charge is spatially confined there was no need for a segmented metallization as required in sub-bandgap excitation setups; therefore a simple pad layout was chosen. The diamond lower pad was glued onto a small PCB and its upper pad was wire-bonded to the input of the amplifier. A CIVIDEC\cite{cividec} C2-HV (2\,GHz, 40\,dB) broadband amplifier/bias-T combination amplified the signals coming from the diamond while providing the bias voltage to the diamond. No active cooling or heating was used. All measurements were performed at room temperature.

The measurements were fully automated. A computer controlled the 3-axis movement of the sample while constantly logging the leakage current and acquiring waveforms from the laser-triggered oscilloscope. The xyz-stage consisted of 3 linear Newport stages with sub-micron resolution driven by a Newport ESP301 motion controller. The 2\,GHz bandwidth, 10\,GS/s sampling rate Tektronix MSO5204B oscilloscope allowed to digitize and record, synchronous to the laser repetition rate, a series of 20 to 100 waveforms with 8-bit pulse height resolution and a length of 200\,ns.

\begin{figure}[ht]
  \center
    \includegraphics[width=0.47\textwidth]{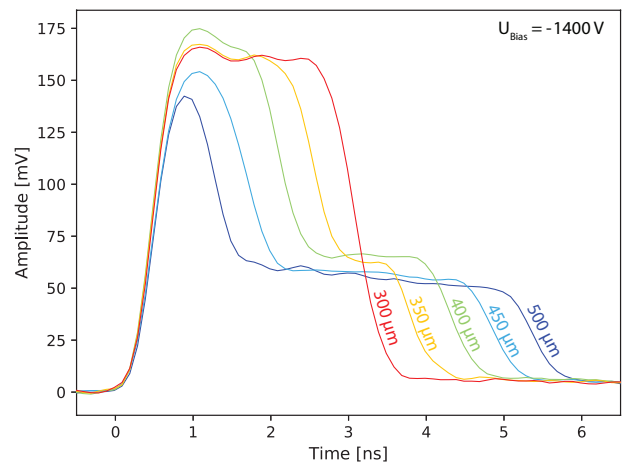}
  \caption{Averaged and baseline corrected waveforms taken at 5 different positions with varying $y$-coordinates and constant {$x$=2500\,$\mu m$} and {$z$=1000\,$\mu m$} coordinates (see figure~\ref{pic:charge2D} for reference). The signals shown are the average of 100 acquisitions with a baseline correction calculated from 1200 waveforms recorded outside the diamond region.}
  \label{pic:waveforms}
\end{figure}

\indent The first step in the analysis was the averaging and baseline correction of the waveforms. The latter was done by taking the average of a large number of waveforms acquired in air outside the diamond and subtracting the result from the average waveform at every scan point. This method allowed us not only to find the baseline of the waveforms but also to remove noise from the laser Pockels cells as well as recurring pickup of the amplifying circuit. Figure~\ref{pic:waveforms} shows a collection of waveforms taken at -1400 V bias voltage on the top electrode at different distances from the bottom electrode (y-positions). The rising edges of the signals indicates the initial drift of charge carriers just after the laser pulse. Thereafter, the initial peak of the signals at $y$-positions (350-500\,$\rm \mu m$) is a result of both electron and hole drift. The tail reaching up to $\sim$5.5\,ns is caused by the electron cloud drifting to the ground electrode. For $y$=300\,$\rm \mu m$ the drift times of electrons and holes to the respective electrodes are roughly the same and no tail is seen.

\begin{figure}[ht]
  \centering
    \includegraphics[width=0.47\textwidth]{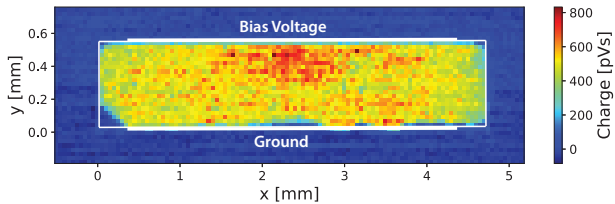}
  \caption{xy-scan of the diamond at z=1000\,$\mu m$ showing collected charge at a bias voltage of {-500\,V}. The thin white lines show the silhouette of the diamond and the two thick lines represent the contact electrodes.}
  \label{pic:charge2D}
\end{figure}

\noindent The collected charge for a scan point at position ($x$,$y$,$z$) can be calculated as
\begin{equation}
Q = \int_{t_a}^{t_b} I(x,y,z,t) dt
\end{equation}

\noindent where $t_a$ and $t_b$ denote the integration times. The integration window was always chosen to integrate the full range of the waveforms (0-200\,ns) in order not to neglect the charges with a smaller drift speed than expected. Performing this procedure for every point with step sizes of $\Delta x$=50\,$\rm \mu m$ and $\Delta y$=25\,$\rm \mu m$ and roughly 4400 measurement points yields a 3-D (x,y at a given z) charge collection map shown in figure~\ref{pic:charge2D}.\par

The ``prompt'' current method\cite{opa_kramberger2010} was used to measure the electric field configuration inside the diamond. Immediately after a laser pulse, electrons and holes that were generated in the voxel of the focal point start drifting in the electric field of the diamond. This initial drift is proportional to the average electric field in the immediate surroundings of the injection voxel. Drifting charges in the diamond cause a current that is measured at the amplifier output. While the drift of the carriers from the injection voxel to the electrodes is characterized by pulse shapes equivalent to the ones shown in figure~\ref{pic:waveforms}, the information about the drift in the first few hundred picoseconds is contained in the rising edge of the current pulses. We measure this initial drift by computing a {200-ps} integral around the center of the rising edge. The integral helps to minimize the uncertainty (e.g. from trigger jitter) in comparison to a single value measurement at the center of the rising edge. The outcome of the integral is called prompt current. The prompt current ($I_{prompt}$) is a measure that only depends on the weighting field $\vec{E}_w$ and the mobilities of electrons and holes near the injection point. This can be shown by using Shockley-Ramo's formula for the instantaneous current\cite{ramo39} multiplied by the amplifier gain {\it A},

\begin{equation}
\begin{split}
I_{promt}(y, t\approx 0) = A q(\vec{v}_e+\vec{v}_h)\vec{E}_w 
\label{ramo}
\end{split}
\end{equation}

\noindent then substituting the definition for mobility $\vec{v}_e+\vec{v}_h = [ \mu_e(E) + \mu_h(E) ] \cdot \vec{E}(y)$ and combining all terms not depending on the electric field {\it E} in a constant {\it c} we obtain: 

\begin{equation}
\begin{split}
I_{promt}(y, t\approx 0) =c [ \mu_e(E) + \mu_h(E) ] \cdot E(y)
\label{prompt_current}
\end{split}
\end{equation}

\begin{figure}[ht!]
  \centering
    \includegraphics[width=0.47\textwidth]{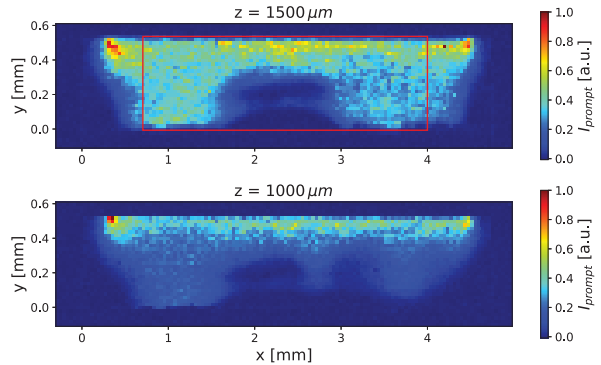}
  \caption{Prompt current xy-maps for {$z=1500\,\mu m$} and  {$z=1000\,\mu m$} at bias voltage of {-500\,V}. The red box in the {$z=1500\,\mu m$} plot is the fiducial area for the comparison between prompt current and X-ray topograph in figure~\ref{pic:agreement}.}
  \label{pic:promptCurrent}
\end{figure}

The strength of this approach is the ability to disentangle the measurement of the electric field from trapping effects that occur during the drift of the charges. Current pulses in figure~\ref{pic:waveforms} have an average 20-80\% rise time of 350\,ps, which at the given field corresponds to an average drift distance of 32\,$\rm \mu m$. The carrier mean free path in electronic grade scCVD diamond is much larger at such electric fields, usually it exceeds the sample thickness. Therefore the trapping during the initial drift can be neglected in this sample. In materials where trapping effects are not negligible compared to the amount of generated charge, equation~\ref{prompt_current} has to be corrected for immediate charge trapping during the initial drift. Figure~\ref{pic:promptCurrent} shows two xy-prompt current maps at two different z-positions in the diamond.  

\begin{figure}[ht!]
  \centering
    \includegraphics[width=0.47\textwidth]{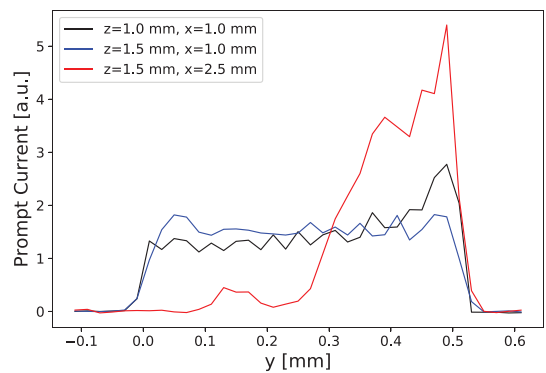}
  \caption{The prompt current profiles for a few selected $x$, $z$ positions as indicated in the legend. For comparison the profiles were normalized to the same area.}
  \label{pic:promptCurrentProj}
\end{figure}

\noindent The plots show strong variations of the prompt current and thus the electric field, which in turn indicates a strong, non-homogeneous variation of space charge in the diamond bulk. In figure~\ref{pic:promptCurrentProj} the prompt current distributions in figure~\ref{pic:promptCurrent} are plotted as a function of $y$ for several selected $x$, $z$ positions. For the profile at {$z=1.5\,mm$} and {$x=2.5\,mm$} the density of space charge is high enough to leave the region at small $y$ without a sizable field. 

\begin{figure}[ht!]
  \centering
    \includegraphics[width=0.45\textwidth]{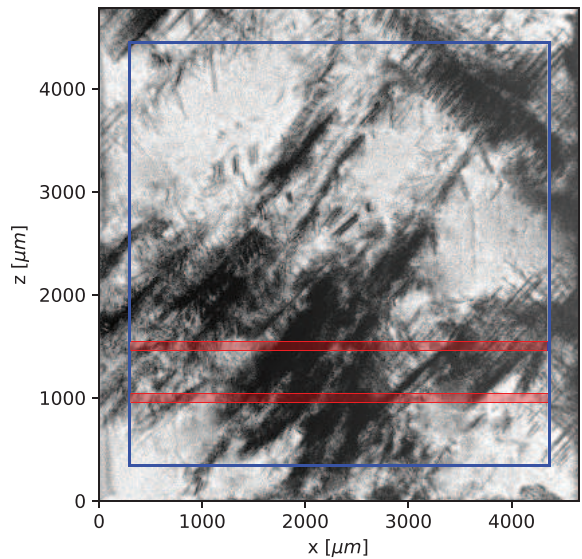}
  \caption{White beam X-ray topograph of the diamond sample. Areas of higher structural defect density appear darker. The outline of the pad metallization is shown in blue. The z-positions of the xy-scans shown in figure \ref{pic:promptCurrent} are indicated in red. The red bars indicate the FWHM-extension of the focal point's squared intensity profile.}
  \label{pic:xraytopo}
\end{figure}

To see whether the variations in the prompt current are caused by structural defects in the diamond, we compared regions of the prompt current xy-scans to slices of an X-ray diffraction topograph taken at the BM05 beamline of the ESRF in Grenoble\cite{esrf}. The (400) Laue spot of the diamond visualizes defects present in the bulk. Dark regions in the topograph in figure~\ref{pic:xraytopo} correspond to a high defect density.

\begin{figure}[ht!]
  \centering
    \includegraphics[width=0.47\textwidth]{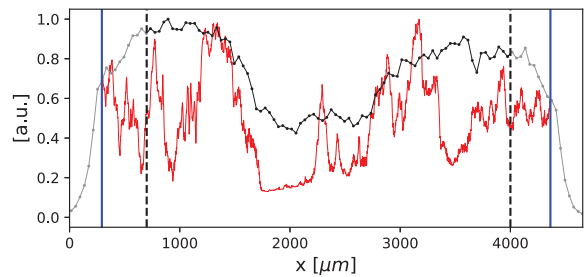}
  \caption{Comparison of the x-projection of the red box in the prompt current map of figure~\ref{pic:promptCurrent} (black) to the focal point's squared intensity profile weighted slice of the X-ray topograph shown as a red bar at z=1500\,$\mu m$ in figure~\ref{pic:xraytopo}. High values in the red curve signifies a high transparency and thus a low defect concentration in the X-ray topograph. The blue lines represent the end of the electrode pad metallization and the dashed black line the fiducial area cut. Data outside the fiducial area cut is shown for completeness.}
  \label{pic:agreement}
\end{figure}

In order to make a comparison between the X-ray topograph and the xy-scan of the prompt current that is not biased by the high field region around the electrode, a fiducial area, shown as red box in figure \ref{pic:promptCurrent}, was selected. The prompt current values were summed for each x-column in the fiducial area, resulting in 1-dimensional distribution in figure~\ref{pic:agreement}. The image histogram of the X-ray topograph was obtained by summing the contrast values of the pixels enclosed by the red bar at z=1500\,$\mu m$ in figure~\ref{pic:xraytopo}. The summation was done along the z-axis for every pixel column along the x-axis. To account for the non-homogenous intensity distribution of the focal point along the z-direction the pixels were weighted with the squared intensity profile of the focal point prior to summation. Both distributions (shown in figure~\ref{pic:agreement}) were normalized to their maximum values for comparison. The electric field drop below 700\,$\mu m$ and above 4000\,$\mu m$ is due to the edge effect. In addition to this, we observe a large drop in the electric field between x $\sim$1500  and x $\sim$ 2800\,$\mu m$. This drop coincides with a large dark region at the same position in the x-ray topograph. The other dark regions in the x-ray topograph do not cause significant disturbance in the electric field, indicating that those defects do not affect the electric field.

Our results complement the X-ray diffraction topograpy. The X-ray diffraction topography indicates the structural defects of the sample. Our method is sensitive to variations of the electric field due to the space charge formed by trapping on defects. It therefore indicates the location and the polarity of the trapped charge. In addition, our method explores the defects in 3 dimensions while the X-ray topograph in our case is 2-dimensional. Figure~\ref{pic:agreement} exhibits a correlation indicating a link between structural defects and charge trapping centers. In our future work we will perform a more systematic study of the correlation between the structural defects and their effect on charge trapping and space charge distribution. In addition to correlation with X-ray topography, it will be enlightening to explore correlations with other defect imaging methods, such as the recently reported photoelectrical readout of single defects\cite{siyushev}.

\indent In summary, we demonstrated the application of the TPA edge-TCT technique in a wide-bandgap semiconductor. We presented 3-D resolved measurements of the charge collection and the electric field in the diamond sample. We have shown that a correlation between the electric field in the diamond and the dislocations observed in the X-ray topography image exists. In addition we have demonstrated that the measurements of the charge collection and the electric field can be achieved with an excellent position resolution in all 3 dimensions. This was previously difficult to achieve in wide-bandgap materials, but is now accomplished through the use of two-photon absorption.

\begin{acknowledgments}
We would like to thank Robert Stone from Rutgers University for providing us with the diamond sample, Marcos Fern\'andez Garc\'ia of the CERN Silicon Lab for many useful discussions that led to these measurements. Special thanks to Professor Ursula Keller at ETH for letting us use her group's laser system and Lukas Gallmann for his consulting on ultrafast laser physics. We would also like to thank Professor Steven Johnson at ETH for providing us with a laser and a space in his lab that allowed us to continue this experiment, and also Martin Kubli and Larissa Boie for their advice and support there. This work was supported by ETH grant ETH-51 15-1. 
\end{acknowledgments}


\begin{thebibliography}{10}

\bibitem{green_diamond}
T.~T. Pham, N.~Rouger, C.~Masante, G.~Chicot, F.~Udrea, D.~Eon, E.~Gheeraert,
  and J.~Pernot, ``Deep depletion concept for diamond {MOSFET},'' {\em Applied
  Physics Letters}, vol.~111, p.~173503, 2017.

\bibitem{Lukas_PSD}
L.~B\"ani, A.~Alexopoulos, M.~Artuso, F.~Bachmair, M.~Bartosik, J.~Beacham, H.~Beck, V.Bellini, V.~Belyaev, B.Bentele {\em et~al.}, ``Diamond detectors for high energy physics
  experiments,'' {\em JINST}, vol.~13, no.~01, p.~C01029, 2018.

\bibitem{gabrysch}
M.~Gabrysch, S.~Majdi, D.~J. Twitchen, and J.~Isberg, ``Electron and hole drift
  velocity in chemical vapor deposition diamond,'' {\em Journal of Applied
  Physics}, vol.~109, no.~6, p.~063719, 2011.

\bibitem{opa_kramberger2010}
G.~Kramberger, V.~Cindro, I.~Mandi\'c, M.~Miku\v{z}, M.~Milovanovi\'c,
  M.~Zavrtanik, and K.~\v{Z}agar, ``Investigation of {I}rradiated {S}ilicon
  {D}etectors by {E}dge-{TCT},'' {\em IEEE Transactions on Nuclear Science},
  vol.~57, pp.~2294--2302, Aug 2010.

\bibitem{tpa2}
F.~R.~Palomo, I.~Vila, M.~Fern\'andez, P.~DeCastro, M.~Moll, ``Two photon absorption and carrier generation in
  semi-conductors.'' 25th RD50 General Meeting, 2014.

\bibitem{tpa3}
I.~Vila, F.~R.~Palomo, M.~Fern\'andez, M.~Moll, P.~DeCastro, R.~Montero, ``A novel transient-current-technique based on the two
  photon absorption process.'' 25th RD50 General Meeting, 2014.

\bibitem{tpa4}
M.~F. Garc\'{i}a, J.~G. S\'{a}nchez, R.~J. Echeverr\'{i}a, M.~Moll, R.~M.
  Santos, D.~Moya, R.~P. Pinto, and I.~Vila, ``High-resolution
  three-dimensional imaging of a depleted {CMOS} sensor using an edge
  {T}ransient {C}urrent {T}echnique based on the {T}wo {P}hoton {A}bsorption
  process ({TPA}-e{TCT}),'' {\em Nuclear Instruments and Methods in Physics
  Research Section A: Accelerators, Spectrometers, Detectors and Associated
  Equipment}, vol.~845, pp.~69 -- 71, 2017.
\newblock Proceedings of the Vienna Conference on Instrumentation 2016.

\bibitem{uflgroup}
See \url{http://www.ulp.ethz.ch} for more information about the used (Attoline)
  laser system; accessed 24-January-2019.

\bibitem{lucchini_laser}
R.~Locher, M.~Lucchini, J.~Herrmann, M.~Sabbar, M.~Weger, A.~Ludwig,
  L.~Castiglioni, M.~Greif, M.~Hengsberger, L.~Gallmann, and U.~Keller,
  ``Versatile attosecond beamline in a two-foci configuration for simultaneous
  time-resolved measurements,'' {\em Review of Scientific Instruments},
  vol.~85, no.~1, p.~013113, 2014.

\bibitem{Kozak12}
M.~Koz\'{a}k, F.~Troj\'{a}nek, B.~Dzur\v{n}\'{a}k, and P.~Mal\'{y}, ``Two- and
  three-photon absorption in chemical vapor deposition diamond,'' {\em J. Opt.
  Soc. Am. B}, vol.~29, pp.~1141--1145, May 2012.

\bibitem{Note1}
Element Six Technologies Limited, Kings Ride Park, Ascot, Berkshire, SL5 8BP,
  United Kingdom.

\bibitem{cividec}
See \url{http://www.cividec.at} for amplifier characteristics; accessed
  24-January-2019.

\bibitem{ramo39}
S.~Ramo, ``Currents {I}nduced by {E}lectron {M}otion,'' {\em Proceedings of the
  IRE}, vol.~27, pp.~584--585, Sept 1939.

\bibitem{esrf}
See \url{https://www.esrf.eu/} for an overview of the facilities at ESRF;
  accessed 24-January-2019.
  
\bibitem{siyushev}
P.~Siyushev, M.~Nesladek, E.~Bourgeois, M.~Gulka, J.~Hruby, T.~Yamamoto, 
M.~Trupke, T.~Teraji, J.~Isoya and F.~Jelezko, ``Photoelectrical imaging and coherent spin-state readout of single nitrogen-vacancy centers in diamond,'' {\em Science}, vol.~363, pp.~728--731, February 2019.

\end{thebibliography}
\end{document}